\documentclass[twocolumn,pra,aps,superscriptaddress,english,floatfix]{revtex4-1}
\usepackage{amsmath}
\usepackage{amssymb}
\usepackage{amsfonts}
\usepackage[pdftex]{graphicx}
\usepackage[dvipsnames]{xcolor}
\usepackage{subfigure} 
\usepackage[english]{babel}
\usepackage{braket}
\usepackage{color}
\usepackage{mathtools}
\usepackage{threeparttable}
\usepackage{ragged2e}
\emergencystretch=\maxdimen
\usepackage[colorlinks,linkcolor=blue,anchorcolor=blue,citecolor=blue,urlcolor=blue]{hyperref}
\usepackage{babel}
\makeatother

\begin{document}

\preprint{APS/123-QED}

\title{Spectral Phase Transitions  in  Optical Parametric Oscillators}% Force line breaks with \\

\author{Arkadev Roy}
\affiliation{%
Department of Electrical Engineering, California Institute of Technology, Pasadena, California 91125, USA}%
%\email{First.Author@caltech.edu} 
\author{Saman Jahani}%
\affiliation{%
Department of Electrical Engineering, California Institute of Technology, Pasadena, California 91125, USA}%
%\email{Second.Author@caltech.edu}
 \author{Carsten Langrock}
 \affiliation{%
Edward L. Ginzton Laboratory, Stanford University, Stanford, California, 94305, USA}%
 \author{Martin Fejer}
 \affiliation{%
Edward L. Ginzton Laboratory, Stanford University, Stanford, California, 94305, USA}%
 \author{Alireza Marandi}
 \email{marandi@caltech.edu}
\affiliation{%
Department of Electrical Engineering, California Institute of Technology, Pasadena, California 91125, USA}%

%\affiliation{
% Second institution and/or address
%
%\affiliation{

\begin{abstract}
Spectral behaviors of photonic resonators have been the basis for a range
of fundamental studies, with applications in classical and quantum technologies \cite{vahala2003optical,limonov2017fano}. Driven nonlinear
resonators provide a fertile ground for phenomena related to  phase transitions  far from equilibrium \cite{cross1993pattern}, which can open opportunities unattainable in their linear counterparts. Here, we show that optical parametric oscillators (OPOs) can undergo second-order phase transitions in the spectral domain between degenerate and non-degenerate regimes. This abrupt change in the spectral response follows a square-root dependence around the critical point, exhibiting high sensitivity to parameter variation akin to systems around an exceptional point \cite{ozdemir2019parity}.  We experimentally demonstrate such a phase transition in a quadratic OPO, map its dynamics to the universal Swift-Hohenberg equation, and extend it to Kerr OPOs. To emphasize the fundamental importance and consequences of this phase transition, we show that the divergent susceptibility of the critical point is accompanied by spontaneous symmetry breaking and distinct phase noise properties in the two regimes, indicating the importance of a beyond nonlinear bifurcation interpretation. We also predict the occurrence of first-order spectral phase transitions in coupled OPOs. Our results on non-equilibrium spectral behaviors can be utilized for enhanced sensing \cite{yang2019quantum,hodaei2017enhanced,lai2019enhanced}, advanced computing \cite{marandi2014network}, and quantum information processing \cite{carusotto2013quantum,gatti1995quantum}.

\end{abstract}

\maketitle

Photonic resonators appearing in myriad forms ranging from macro-scale to nano-scale have been the mainstay of light-based fundamental studies and applications \cite{vahala2003optical}. The ability to engineer the resonant spectral features of these cavities unveil tremendous possibilities in sensing and light-matter interactions. The interplay of gain/loss and coupling in coupled linear photonic resonators can lead to the occurrence of a multitude of intriguing phenomena ranging from Fano resonance, electro-magnetically induced transparency, \cite{limonov2017fano} and exceptional point associated with parity-time symmetry breaking \cite{ozdemir2019parity, miri2019exceptional}. 

Strong nonlinearities in photonic resonators can lead to a variety of rich phenomena. Nonlinear driven dissipative systems existing in non-equilibrium steady states exhibit self-organization \cite{ropp2018dissipative}, pattern formation
\cite{vaupel1999observation, bortolozzo2001experimental,lugiato1987spatial,taranenko1998pattern,oppo2013self}, and emergent phase and dynamical phase transitions \cite{cross1993pattern}. Other salient examples include behaviors in laser systems \cite{kuznetsov1991optical, marowsky1978second, haken1975cooperative} at threshold \cite{degiorgio1970analogy} and around mode-locking transitions \cite{leonetti2011mode,gordon2002phase,fischer2013many}, soliton-steps in
Kerr micro-resonators \cite{herr2014temporal}, and in polaritonic quantum fluids \cite{ carusotto2013quantum}. Similar phenomena are also explored outside photonics for instance in the form of Rayleigh-Benard convection and Faraday waves \cite{bodenschatz2000recent, engels2007observation}. 

Specific to the parametric oscillation regime, a variety of nonlinear dynamical behaviors has been predicted and demonstrated such as bi-stability, self-pulsation, limit-cycles \cite{drummond1980non}, pattern formation \cite{vaupel1999observation, bortolozzo2001experimental,lugiato1987spatial,taranenko1998pattern,oppo2013self} and phase transitions \cite{dunnett2016keldysh, dagvadorj2015nonequilibrium}, albeit not explicitly in the spectral domain. Here, we exploit the rich dynamics of nonlinear driven dissipative systems in OPOs to formulate and engineer their spectral behaviors as phase transitions.

Phase transition marks a universal qualitative regime change in system properties as the control parameter is varied around a critical/transition point \cite{stanley1971phase}.  The behavior of the system around the critical point is characterized by the order parameter (OP). Second-order phase transition displays continuity in the OP while exhibiting a discontinuity in the derivative of the OP. On the other hand, first-order transition is known to possess a discontinuous OP around the transition point.  \\

Realizing phase transitions based on the optical parametric processes can provide unique opportunities for sensing. For instance, in phase-transition-based detectors and transition-edge sensors \cite{gol2001picosecond}, the reset time (return time to the critical bias point) can be significantly reduced using an ultrafast nonlinear process compared to thermodynamic transitions. Moreover, similar to the exceptional points in optical systems \cite{miri2019exceptional, ozdemir2019parity}, an enhanced sensitivity \cite{hodaei2017enhanced, lai2019enhanced} can be realized using a driven dissipative-based spectral phase transition. However, in contrast to exceptional points in PT-symmetric systems, this enhancement is not accompanied by eigenvectors non-orthogonality and can potentially provide high sensitivity combined with high precision \cite{wang2019petermann, lau2018fundamental}. The noiseless nature of parametric amplification \cite{caves1982quantum} can be another unique resource for enhancing the signal-to-noise ratio; a property that is not readily available in current implementations of exceptional points.  Divergent susceptibility of the critical point supported by the parametric gain in a driven-dissipative setting can open unexplored avenues in the context of non-Hermitian sensing. 

Spectral phase transitions in OPOs can also open opportunities for computing and quantum information processing. Phase transition occurring at the oscillation threshold of OPOs has been utilized as a promising computing resource in optical Ising machines  \cite{marandi2014network, hamerly2019experimental}. Phase transition occurring in the spectral domain can provide additional computing capabilities in them or enable similar non-Von Neumann computing architectures operating in the spectral domain.

In this work, we consider a doubly-resonant OPO \cite{hamerly2016reduced, trillo2001parametric} as a driven-dissipative system in a non-equilibrium steady state. The driving is accomplished by the synchronous pulsed pump centered around the frequency $2\omega_{0}$, while the resonant signal and the idler constitute the longitudinal modes of the resonator centered around the half-harmonic frequency ($\omega_{0}$).The interaction among the modes is engendered by the quadratic non-linearity (Fig. 1a).  The inherent coupled nature of the signal and idler in a doubly-resonant OPO gives rise to rich nonlinear dynamics including the appearance of bi-phase states around degeneracy \cite{marandi2014network}. The mutual interplay between the cavity detuning and the temporal group velocity dispersion provides another degree of freedom, which governs the dynamics of signal/idler in synchronously pumped doubly resonant OPOs. This leads to discontinuities typical of a second-order phase transition around the critical cavity detuning (Fig. 1b and 1c). This spectral phase transition demarcates the sharp boundary between the degenerate and non-degenerate parametric oscillation.    

\begin{figure}[!h]
\centering
\includegraphics[width=0.45\textwidth]{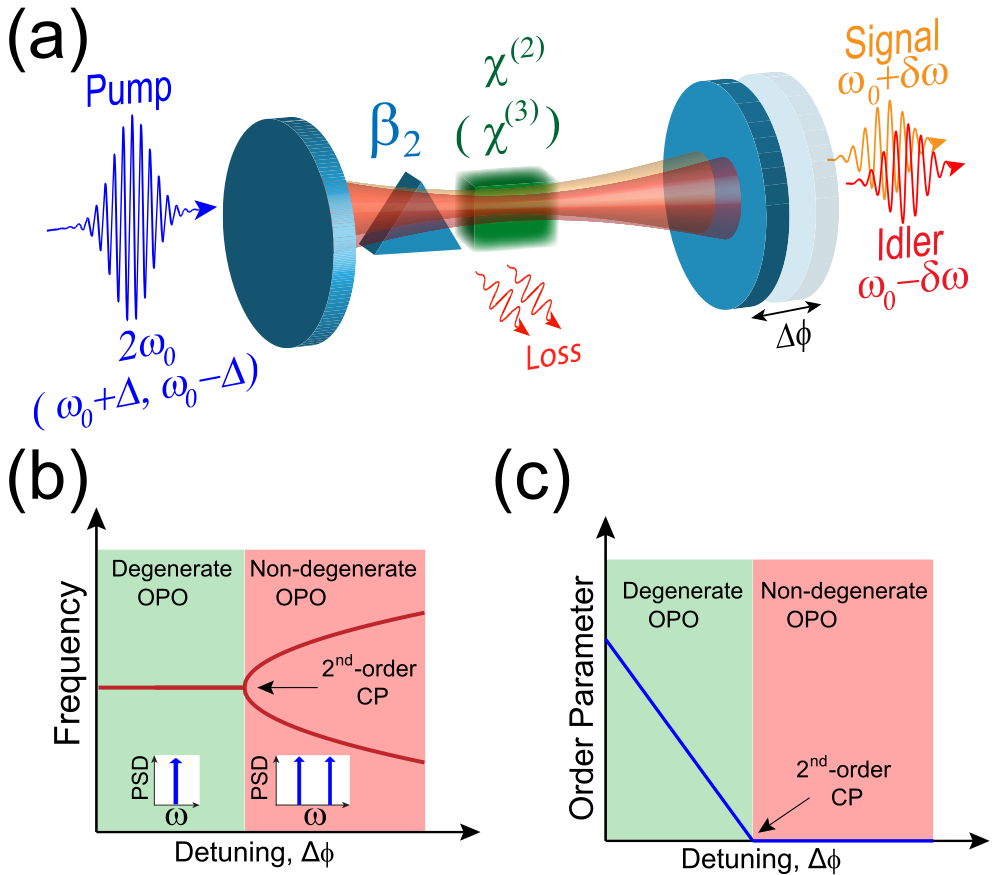}
\caption{\label{fig: schematic} \textbf{Spectral phase transition in nonlinear photonic resonators}. a) OPO with the resonant signal and idler in the cavity with variable detuning $(\Delta\phi)$ and second-order group-velocity dispersion $(\beta_{2})$. The nonlinearity can be provided by a quadratic $(\chi ^{\left(2 \right) })$ or a Kerr $(\chi^{\left( 3\right)})$ medium. b) A second-order phase transition occurs at the critical detuning that marks the transition between the degenerate and the non-degenerate spectrum. c) This transition is characterized by a continuous order parameter, but a discontinuous derivative of the order parameter at the critical point (CP).}
\end{figure}

 In the CW-driven high-finesse limit, the OPO is governed by the mean-field evolution equation:
\begin{eqnarray}
    \frac{\partial a}{\partial \xi} &=& (-\alpha + i\Delta\phi )a + ga^{*} -i\frac{\beta _{2}}{2}\frac{\partial ^{2}a}{\partial t^{2}} \nonumber \\
   & & - \left [ \frac{\epsilon ^{2}}{2u^{2}}  \int_{0}^{Lu} (Lu- \tau )a(t-\tau )^{2}  d\tau\right ] a^{*},
\end{eqnarray}
where, $a$ describes the signal envelope under the slowly varying envelope approximation limit.  Here $\xi$, $t$ refers to the slow time and the fast time respectively \cite{coen2013modeling}. $\alpha$,  $\Delta\phi$, $\beta_{2}$, and $g$ denotes the loss, detuning, group-velocity dispersion (GVD), and the phase-sensitive parametric gain respectively. $g$ in the CW-limit is expressed as $\epsilon b L$, where $b$ is the pump amplitude. $L$ refers to the cavity round trip length where the nonlinear interaction is encountered, $\epsilon$ includes the strength of the nonlinear interaction and $u$ is the walk-off parameter. The last term to the right of the equation is responsible for the gain saturation. Each of these terms are normalized by suitable normalization factors (see Supplemental Section 3). 
 
We assume a perturbation in the field (signal/idler) around the half-harmonic frequency ($\omega_{0}$) to be of the form: $a= a_{+}e^{i\delta\omega t} + a_{-}e^{-i\delta\omega t}$. We perform linear stability analysis (neglecting gain saturation) to determine the most unstable longitudinal mode, which is given as:
\begin{subequations}
\label{eq:whole}
 \begin{gather}
 \frac{d}{d\xi}  \begin{bmatrix} {a_{+}}  \\ {a_{-}^{*}}  \end{bmatrix}
 = \begin{bmatrix}
  -\alpha+i\Gamma &
   g \\
   g^{*} &
   -\alpha-i\Gamma
   \end{bmatrix} 
   \begin{bmatrix} {a_{+}}  \\ {a_{-}^{*}}  \end{bmatrix}
\label{subeq:1}
\end{gather}
\begin{equation}
\lambda _{\pm} =-\alpha \pm \sqrt{|g|^{2}-\Gamma^{2}}
   \label{subeq:2}
\end{equation}
\end{subequations}
where $\Gamma = \Delta\phi +\frac{\beta _{2}}{2} (\delta\omega) ^{2}$.
Analyzing the eigenvalue (growth rate) (Eq(2b)) of the linear stability matrix we arrive at two scenarios. First, when sgn$(\Delta\phi)=$sgn$(\beta_{2})$, we find that the most unstable frequency of oscillation is $\delta\omega=0$, and the corresponding threshold (i.e. when $\lambda_{+}=0$) is $|g|_{th}=\sqrt{\alpha^{2}+(\Delta\phi)^{2}}$, leaving the OPO in the degenerate phase. However, when sgn$(\Delta\phi)=-$sgn$(\beta_{2})$, the most unstable frequency of oscillation is given by $\delta\omega=|\frac{2\Delta\phi}{\beta_{2}}|^{\frac{1}{2}}$, and the associated threshold  is $|g|_{th}=\alpha$, leaving the OPO in the non-degenerate phase. This can be understood as cavity detuning $(\Delta \phi)$ counterbalancing the GVD induced detuning in the non-degenerate regime. This can happen for positive cavity detuning in the anomalous regime, where GVD induced detuning is negative and they cancel exactly at $\omega_{0} \pm \delta\omega$, thereby experiencing more gain in the non-degenerate phase resulting in OPO selecting non-degeneracy over degenerate oscillation. This proves the existence of the spectral phase transition which is demonstrated in Fig 2. The spectral phase transition takes place around the detuning, $\Delta\phi=0$. The behavior in the normal GVD regime (Fig. 2d) is reversed as compared to the anomalous GVD scenario (Fig. 2b). Results obtained experimentally (Fig. 2c, Fig 2.e) agree  well with the simulation. 

The spectral transition can be interpreted as an order-disorder transition whereby the OPO transits from an ordered bi-phase state in the degenerate regime, to a dis-ordered phase state in the non-degenerate regime with the signal assuming random phases and the idler following it (see Supplemental Section 15) \cite{SI}. Thus the critical point marks the onset of the spontaneous U(1) symmetry breaking. In our context we define OP as, $OP = \frac{d \lambda_{max}}{d \Delta\phi} $, which represents the derivative of the gain with respect to the detuning. The gain $( \lambda_{max}$ = ${\displaystyle\max_{\delta\omega}\lambda_{+}} )$ i.e. the maximum eigenvalue is obtained using Eq(S.8) (see Supplemental Section 4).  The phase-dependent parametric gain is sensitive to detuning induced phase accumulation more acutely in the degenerate regime as opposed to the non-degenerate regime where it varies slowly with detuning. The order-disorder transition has important implications in the phase noise and coherence properties of the OPO. While the phase noise of OPO operating at degeneracy is dominated by the driving pump, in the non-degenerate regime phase diffusion leads to Schawlow- Townes limit for each of the signal and the idler \cite{nabors1990coherence}, albeit with anti-correlation in their phases and potential phase-sum quadrature squeezing \cite{fabre1989noise}.  The phase transition description reveals interesting correlation properties in the dis-ordered regime i.e. the non-degenerate regime. The phase difference diffusion follows a power-law dependence as a function of detuning (i.e. distance from the critical point) which mimics the behavior of correlation functions in continuous phase transitions (see Supplemental Section 9).   
\begin{figure}[!h]
\centering
\includegraphics[width=0.5\textwidth]{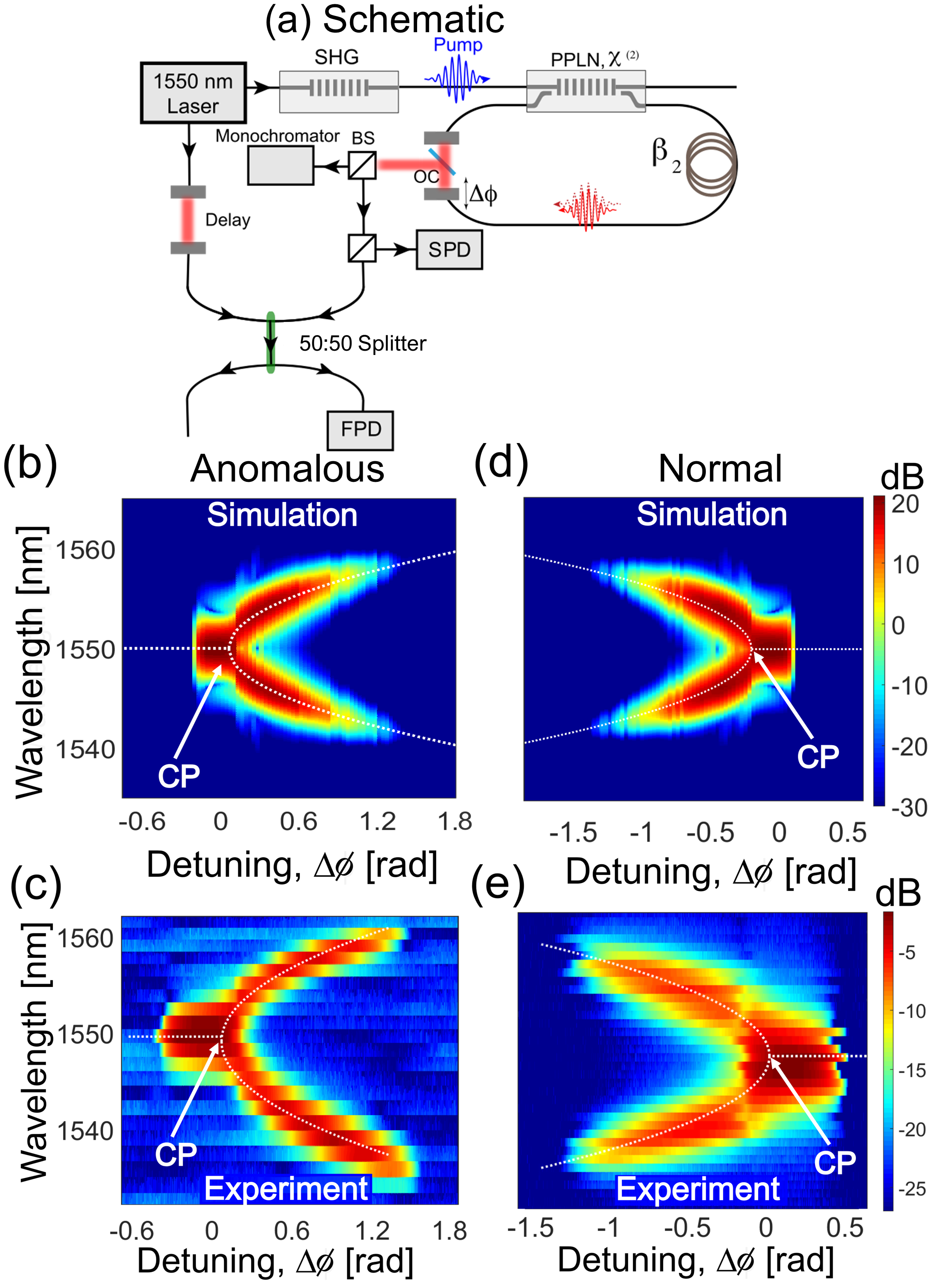}
\caption{\label{fig: Fig2} \textbf{Second-order spectral phase transition in an OPO}. a) Schematic of the experimental setup, which implements the spectral phase transition in a guided-wave OPO based on PPLN. Beam Splitter (BS), Output Coupler (OC), Slow Photo-detector (SPD), Fast Photo-detector (FPD), Second Harmonic Generation (SHG).  Spectrum as a function of detuning obtained through numerical simulation b) in anomalous dispersion regime (dotted line plots the theoretically expected spectral splitting, which in the non-degenerate regime is given by: $\delta \omega = \sqrt{\frac{-2\Delta\phi}{\beta_{2}}}$, d) in the normal dispersion regime.  Experimental results capturing the second-order critical point c) in anomalous dispersion regime, e) in normal dispersion regime. It closely follows the square-root behavior (dotted line) in the non-degenerate regime. Colorbar represents power spectral density in dB.   }
\end{figure}

We further characterize the quadratic OPO around the phase transition point (Fig. 3). The critical point coincides with the maximum output power of the OPO as observed numerically and experimentally (Fig. 3a and 3b). This behavior can be explained by the gain calculations (inset of Fig. 3c). The threshold is a function of detuning and dispersion \cite{SI}. The order parameter displays characteristics (Fig. 3c) typical of second-order phase transitions or soft transitions \cite{strogatz2018nonlinear}.  Additionally, in the pulsed regime as the OPO undergoes the phase transition the signal and idler combs  split and interfere with each other with a beat frequency equal to the difference of their respective carrier-envelope offset frequencies. This leads to the spontaneous emergence of beat notes as shown in the measurement results of Fig. 3d. This is a manifestation of a critical slowing down phenomenon, where the time period of the beat-note tends to infinity as we approach the critical point from the non-degenerate regime.  Note that, the detuning range of the parametric oscillation, as well as the ratio of degenerate and non-degenerate regimes above the threshold is determined by the gain, which is a function of the pump power and cavity dispersion (see Supplemental Section 11). % and experimentally (Fig 3.d). 
\begin{figure}[!h]
\centering
\includegraphics[width=0.5\textwidth]{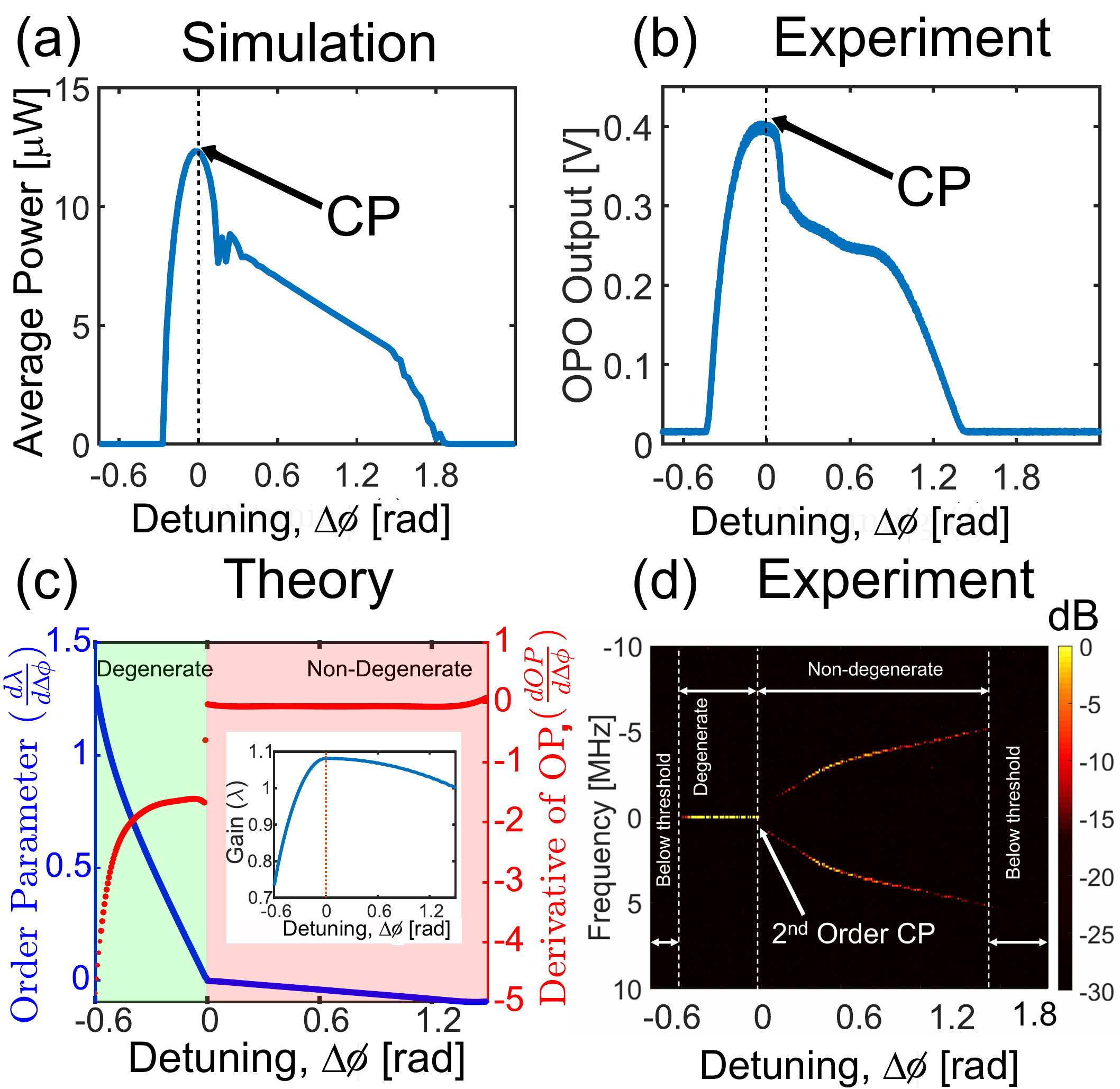}
%\justifying
\caption{\label{fig: Fig3} \textbf{Characterization of the second-order critical point.}  OPO average output power as a function of cavity detuning obtained numerically (a) and experimentally (b) using a slow detector. This demonstrates the maximum conversion efficiency at the critical point. c) continuous order parameter but discontinuous derivative typical of a second order phase transition. The inset shows the gain curve as a function of cavity detuning, which has its maximum at the critical point. d) Spontaneous emergence of beat-note around the critical point. Measured RF spectrum captured using a fast detector in a multi-heterodyne measurement showing co-existence of the signal and idler combs in the non-degenerate regime and their offset tuning.}
\end{figure} 

 When two OPOs are coupled, the transition from degenerate to non-degenerate operation can occur as a first-order phase transition. Fig 4.a depicts a schematic representation of the coupled OPO configuration. In the presence of the coupling, the competition between the two second-order phase transitions (as shown by the gain curve in Fig. 4d) results in the emergence of a first order spectral phase transition (Fig. 4e). This first-order transition point causes a sudden discontinuity/hard transition in the spectrum (Fig. 4b, Fig. 4c) as the coupled OPO  transits from the non-degenerate to the degenerate spectral regime (Fig. 4e). The coupling in the linear regime induces a mode splitting which is expected to introduce a second-order phase transition around the split resonances as evident from the plot of OP in Fig. 4d. This can be understood by the argument that a positive cavity detuning applied to individual cavities can appear both as a positive or a negative detuning in the coupled basis depending on the relative magnitude of the cavity detuning and the coupling strength. Further details regarding the modeling of coupled OPOs is presented in the supplementary information (see Supplemental Section 6,7).
\begin{figure}[!h]
\centering
\includegraphics[width=0.5\textwidth]{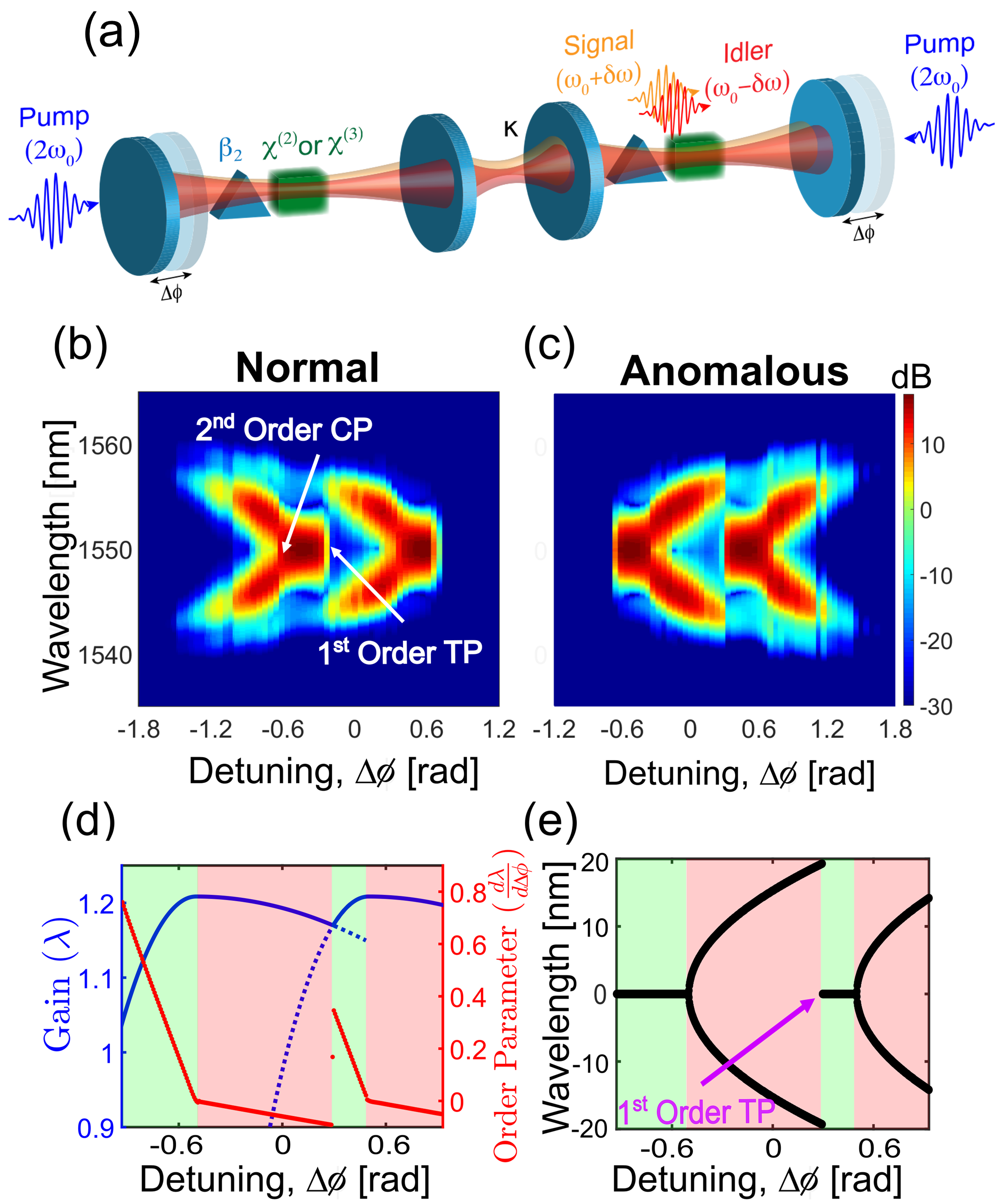}
%\justifying
\caption{\label{fig: Fig4} \textbf{First-order spectral phase transition in coupled OPOs.} a) Schematic configuration of a coupled OPO (coupling factor ($\kappa$)). b,c) Simulated spectrum as a function of cavity detuning b) in the normal dispersion regime, c) in the anomalous dispersion regime. d) Order parameter as a function of detuning showing the discontinuity at the location of the first-order transition point. The gain curve is also plotted alongside. The OPO selects the gain maximum and therefore doesn't follow the dashed portion of the gain curve. This gain competition between two second-order critical point gives rise to the first order transition point. e) The spectrum considering the wavelength experiencing the maximum gain around which the signal/idler is centered. At the first-order transition, there is a discontinuous jump from the non-degenerate spectrum to the degenerate spectrum.  }\label{fig4}
\end{figure}  

The demonstrated spectral phase transitions can be described by the universal Swift-Hohenberg equation which is also known to govern nonlinear pattern formation dynamics \cite{de1996transverse, longhi1996swift}. The mapping of the OPO dynamics to the Swift-Hohenberg equation is derived in S8 (see Supplemental Section 8) \cite{SI}. The same equation we derived in this context can describe degenerate four-wave mixing dynamics contingent to certain conditions. Thus, spectral phase transitions are also expected to occur in Kerr OPOs (Fig. 5a).
\begin{figure}[!h]
\centering
\includegraphics[width=0.5\textwidth]{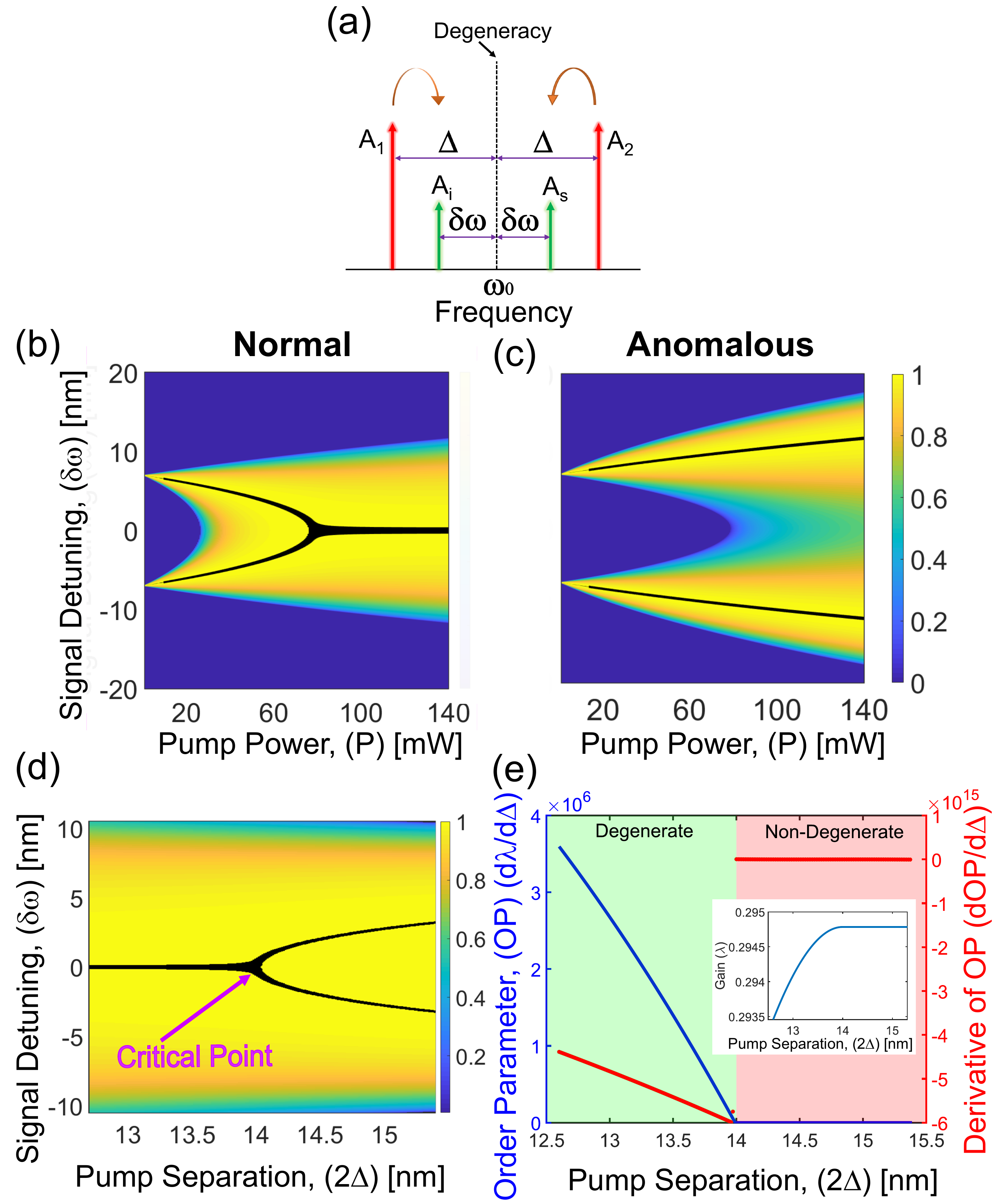}
%\justifying
\caption{\label{fig: Fig4} \textbf{Spectral phase transition in dual-pump four-wave mixing.} a) Illustration of the two pump fields getting converted to the signal and idler. b) The phase conjugation parametric instability gain (Eq 3) curve in the normal dispersion regime as the pump power is varied. The maximum of the gain where the signal/idler is supposed to oscillate is marked by the black lines. Degenerate and non-degenerate oscillations are both expected in this case. c) The phase conjugation parametric process gain curve in the anomalous dispersion regime. Only non-degenerate oscillation is expected in this case. d) spectral phase transition (normal dispersion regime) as the pump separation ($\Delta$) is varied. A degenerate to the non-degenerate transition happens across the critical point. e) The critical point is characterized to be a second-order which displays continuous behavior in order-parameter but exhibits discontinuity in its derivative. Parameters used in the simulation are taken from \cite{okawachi2015dual}.  }
\end{figure} 

For the Kerr OPO, we consider a conservative system governed by the nonlinear Schroedinger equation.  $\frac{\partial A}{\partial z}=-i\frac{\beta_{2}}{2}\frac{\partial^{2} A}{\partial \tau^{2}}+i\gamma|A|^{2}A$. $\gamma$ represents the effective third-order nonlinear co-efficient and $\beta_{2}$ stands for the second order GVD co-efficient.  Degenerate parametric oscillation can be realized in a Kerr medium using dual pumps (Fig. 5b and 5c) \cite{okawachi2015dual, inagaki2016large}. We represent the dual pumps as having amplitudes $A_{1}$ and $A_{2}$ and assume that they have equal power ($P=|A_{1}|^{2}=|A_{2}|^{2}$) and possesses a detuning of $\Delta$ from the center of degeneracy. Owing to the symmetry, we assume the parametrically generated signal ($A_{s}$) and idler ($A_{i}$) to be detuned by $\delta\omega$ from the center of degeneracy. We express the field as given by the following expansion: $A(z,\tau)= A_{1}e^{i\Delta \tau} + A_{2}e^{-i\Delta \tau}+A_{s}(z)e^{i\delta\omega\tau}+A_{i}(z)e^{-i\delta\omega \tau}$. Parametric gain at the onset of the phase conjugation parametric process can be determined via a linear stability analysis. The growth rate due to phase conjugation parametric process can be expressed as $e^{\lambda z}$\, where $\lambda$ is given by \cite{okawachi2015dual}: 
\begin{equation}
 \lambda = \sqrt{\left\{
 6\gamma P - \beta _{2}(\Delta^{2}-(\delta\omega)^{2})\right\}\left\{
 2\gamma P + \beta _{2}(\Delta^{2}-(\delta\omega)^{2})\right\}}  
\end{equation} \\
The spectral phase transition is shown in Fig. 5d. The fact that the associated critical point is second-order is established by analyzing the OP as depicted in Fig. 5e.

The abrupt frequency splitting around the critical point in these spectral phase transitions can be utilized for enhanced sensing. A sensor can be based on the second-order spectral phase transition biased at the critical point, that will exhibit a scaling of $\delta \omega \sim \varepsilon^{\frac{1}{2}}$, where $\varepsilon$ is the small perturbation (e.g. in detuning) under consideration, similar to a second-order exceptional point \cite{hodaei2017enhanced}. However, if we leverage the first-order spectral phase transition for a critical detector, we can utilize the discontinuity in the spectrum for highly enhanced sensitivity \cite{gol2001picosecond,yang2019quantum}. The proportionality constant in the scaling law is a function of the cavity group-velocity dispersion. The smaller the dispersion, higher is the sensitivity (see Supplemental Section 5) \cite{SI}. The presented spectral phase transition can also be utilized in computing architectures. For example, in the OPO-based Ising machines, which have been strictly operating at degeneracy so far \cite{marandi2014network, hamerly2019experimental}, the spectral phase transition can act as an additional search mechanism leveraging the symmetry breaking and additional phase noise in the non-degenerate regime. Moreover, our results on spectral phase transition can lay the foundation for novel types of phase-transition-based computing platforms \cite{kalinin2020polaritonic}.

Tuning the spectrum of parametric oscillation between degeneracy and non-degeneracy is a well-known concept, and the same is achieved by manipulating the phase matching curve via temperature, voltage control, etc. \cite{eckardt1991optical}. Distinctively, the presented spectral phase transition occurs as a multi-mode co-operative phenomenon \cite{haken1975cooperative} triggered by cavity phase detuning, where dispersion plays a crucial role, while the phase-matching enabled by the periodically poled waveguide remains unaltered.

The presented spectral phase transition is in sharp contrast to intensity-dependent bifurcation ubiquitous in nonlinear optical systems. The spectral bifurcation doesn't arise due to the gain saturation induced nonlinearity (see Supplemental Section 9)\cite{SI}. This is also corroborated by the existence of the quantum image of this above-threshold phenomenon below threshold (where gain saturation is absent) (see Supplemental Section 10) which is consistent with the theoretical predictions in the spatial domain \cite{gatti1995quantum}.
%\begin{table}[!h]
%\begin{tabular}{c}

%\includegraphics[width=0.3\textwidth, height=0.3\textwidth]{Ewins_fig.png}\\

 %\end{tabular} 
%\justify
%\captionof{FIG 6:}{\hspace{6} Spectral phase transition in locked configuration. To be replaced with %Edwins plot. }
%\end{table} 

In summary, we have performed complete characterization of this second-order phase transition both in the temporal and in the spectral (optical and radio-frequency) domain. Experimental results are backed by numerical simulations of the underlying spatio-temporal phenomena and corroborated using mean-field analytical descriptions. We have shown that some of the non-equilibrium spectral behaviors of OPOs can be formulated as a universal phase transition.\

The semi-classical spectral phase transition considered in this work can be extended to the quantum regime below threshold opening a path toward a quantum phase transition in the spectral domain.  The ability to engineer the dispersion of integrated $\chi^{\left (2 \right)}$ and $\chi^{\left( 3\right)}$ devices \cite{wang2018nanophotonic} will allow manipulation of the spectral phase transition behavior. Probing the performance difference of sensors based on second-order phase transitions and those leveraging second-order exceptional points \cite{hodaei2017enhanced, lai2019enhanced} is a subject of future work. \

\begin{acknowledgments}
We  acknowledge stimulating discussions with Avik Dutt, Mohammad-Ali Miri, Myoung-Gyun Suh, Li-Ping Yang, Marc Jankowski. The authors gratefully acknowledge support from ARO Grant No. W911NF-18-1-0285 and NSF Grant No. 1846273 and 1918549, and AFOSR award FA9550-20-1-0040. The authors wish to thank NTT Research for their financial and technical support. \\
\end{acknowledgments}

%\section*{Competing Interests}
%The authors declare no competing interests. \\

%\section*{Author Contribution}
%A.R. and A.M. conceived the idea and performed the experiments. A.R. and S.J. developed the theory and performed the numerical simulations. C.L. fabricated the PPLN waveguide used in the experiment with supervision of M.F. \hspace{10} A.M. supervised the project. A.R. wrote the manuscript with input from all authors.

\section{Experimental Setup}
 The experimental schematic is shown in Fig 2.a, a detailed version of which is presented as Fig S.1 (Supplemental Section 2) \cite{SI}. The OPO pump is derived from the mode-locked laser through second harmonic generation (SHG) in a quasi-phase matched periodically poled lithium niobate (PPLN) crystal. The pump is centered around 775 nm. The main cavity is composed of a PPLN waveguide (reverse proton exchange, 40 mm long, periodically poled to phase-match 775-1550 nm interaction) \cite{langrock2007fiber} with fiber coupled output ports, fiber phase shifter, free-space section (to adjust the pump repetition rate to be multiple of the free spectral range of the cavity.), additional fiber segment to engineer the cavity dispersion, and a beam splitter which provides the output coupling. All fibers and devices existing in the optical path are single mode, polarization maintaining and connectors are angle polished. Additional details pertaining to the experimental setup/methods is provided in the supplementary information (Supplemental Section 2) \cite{SI}.   \
\section{System Modeling}
 The nonlinear interaction inside the PPLN waveguide is governed by: 
\begin{subequations}
\label{eq:whole1}
\begin{equation}
\frac{\partial a}{\partial z} =\left[ - \frac{\alpha^{(a)}}{2}  -i\frac{\beta_{2}^{(a)}}{2!}\frac{\partial^{2}}{\partial t^{2}}+ \ldots\right]a +\epsilon a^{*}b
\label{subeq:3}
\end{equation}
\begin{equation}
\frac{\partial b}{\partial z}=\left[- \frac{\alpha^{(b)}}{2} - u\frac{\partial}{\partial t} -i\frac{\beta_{2}^{(b)}}{2!}\frac{\partial^{2}}{\partial t^{2}} + \ldots \right]b -\frac{\epsilon a^{2}}{2}
\label{subeq:4}
\end{equation}
\end{subequations}

 The evolution of the signal($a$) and  the pump($b$) envelopes in the slowly varying envelope approximation are dictated by (4a) and (4b) respectively \cite{hamerly2016reduced}. The effects of higher-order group velocity dispersions (GVD) $\beta_{2}, \beta_{3} $, group velocity mismatch (GVM) ($u$), the back-conversion from the signal to the pump are included. The round-trip feedback is given by:
 \begin{subequations}
\label{eq:whole2}
\begin{equation}
a^{(n+1)}(0,t) = {\cal F}^{-1}\left\{ G_{0}^{-\frac{1}{2}} e^{i\bar{\phi}} {\cal F} \left\{ a^{(n)}(L,t)\right\}\right\} %
\label{subeq:5}
\end{equation} 
\begin{equation}
\bar{\phi} = \Delta\phi  + \frac{l \lambda^{(a)}}{2c}(\omega-\omega_{0})  + \frac{\phi_{2}}{2!}(\omega-\omega_{0}) ^{2}+ \ldots
\label{subeq:6}
\end{equation}
\end{subequations}

 Eq(5) takes into consideration the round-trip loss which is lumped into a aggregated out-coupling loss factor $G_{0}$, the GVD ($\phi_{2}$) of the feedback path and the detuning ($\Delta\phi$) ($\Delta\phi=\pi l + \phi_{0}$, $l$ is the cavity length detuning in units of signal half-wavelengths in vacuum) of the circulating signal from the exact synchrony with respect to the pump. The effective second-order nonlinearity co-efficient ($\epsilon$) is related to the SHG efficiency \cite{hamerly2016reduced}.  The round-trip number is denoted by $n$ and the cavity length by $L$. The equations are numerically solved adopting the split-step Fourier algorithm. \ 

To explain the spectral phase transition phenomenon numerically and analytically we adopt a two-pronged approach. First, we develop a mean-field model in the high finesse, CW driven limit and provide an analytical description for the occurrence of the spectral phase transition. This model, though does not inherit all the characteristics of the synchronously pumped optical parametric oscillator, is able to encapsulate the qualitative nature of the spectral phase transition. 
Secondly, we calculate the eigenvalue which is related to the gain of the signal/idler per roundtrip by assuming the pump in the OPO to be effectively CW having an average power being equal to the peak power of the pump pulse (see Supplemental Section 4) \cite{SI}.

For Kerr OPO, the  evolution  of  the  optical  fields  in  the  non-resonant  Kerr  nonlinear  medium  is  governed  by  the nonlinear Schrodinger equation. The larger eigenvalue ( $\lambda$) of the linear stability matrix of the phase conjugation nonlinear interaction determines the gain and is obtained in the undepleted pump approximation (see Supplemental Section 14) \cite{SI}. However, in a Kerr nonlinear medium additional nonlinear interactions, namely the modulation instability and four-wave mixing Bragg scattering accompany the phase conjugation process responsible for the phase-sensitive degenerate parametric oscillation. Spectral phase transitions can also be investigated  in a  driven-dissipative  setting  in  a  Kerr  resonator,  with the  detuning  between  the  pumps  and  the  cold-cavity resonances  being  an  additional  degree  of  freedom.  

\nocite{*}
\bibliography{cr}% Produces the bibliography via BibTeX.

\end{document}